\def\comment#1{}

\newcommand{\beg}{\begin{eqnarray}}
\newcommand{\eee}{\end{eqnarray}}

\documentclass[prl,letterpaper,twocolumn,aps,epsf]{revtex4}
\usepackage{dcolumn}
\usepackage{amsmath}
\usepackage{graphicx}
\def\cm#1{}

\begin{document}
\title{ 
Neither a type-I nor a type-II superconductivity in a two-gap system}
\author{
Egor Babaev 
}
\address{
     Institute for Theoretical Physics, Uppsala University, Box 803, 75108 Uppsala, Sweden \\
Cornell University 620 Clark Hall Ithaca, NY 14853-2501 USA
}
\begin{abstract}
We show that in  two-gap Ginzburg-Landau model
there is a wide  range of parameters where  behaviour 
of a superconductor in external field can not 
be classified nor as type-I nor as type-II. In this regime a superconductor features a nonmonotonic
interaction between vortices (which is 
attractive at a certain large length scale and repulsive at a shorter
length scale). This results in a first order transition  into 
an inhomogeneous state of a mixture of vortex clusters 
where one of the order parameters is suppressed
and domains of two-gap Meissner state. 
\end{abstract}
\maketitle
\newcommand{\la}{\label}
\newcommand{\aaa}{\frac{2 e}{\hbar c}}
\newcommand{\Pfaff}{{\rm\, Pfaff}}
\newcommand{\kA}{{\tilde A}}
\newcommand{\G}{{\cal G}}
\newcommand{\cP}{{\cal P}}
\newcommand{\M}{{\cal M}}
\newcommand{\E}{{\cal E}}
\newcommand{\btd}{{\bigtriangledown}}
\newcommand{\W}{{\cal W}}
\newcommand{\X}{{\cal X}}
\renewcommand{\O}{{\cal O}}
\renewcommand{\d}{{\rm\, d}}
\newcommand{\bfi}{{\bf i}}
\newcommand{\e}{{\rm\, e}}
\newcommand{\bfx}{{\bf \vec x}}
\newcommand{\bfn}{{\bf \vec n}}
\newcommand{\bfE}{{\bf \vec E}}
\newcommand{\bfB}{{\bf \vec B}}
\newcommand{\bfv}{{\bf \vec v}}
\newcommand{\bfU}{{\bf \vec U}}
\newcommand{\ccc}{{\vec{\sf C}}}
\newcommand{\bfp}{{\bf \vec p}}
\newcommand{\f}{\frac}
\newcommand{\bfA}{{\bf \vec A}}
\newcommand{\non}{\nonumber}
\newcommand{\be}{\begin{equation}}
\newcommand{\ee}{\end{equation}}
\newcommand{\ba}{\begin{eqnarray}}
\newcommand{\ea}{\end{eqnarray}}
\newcommand{\bastar}{\begin{eqnarray*}}
\newcommand{\eastar}{\end{eqnarray*}}
\newcommand{\half}{{1 \over 2}}
\newcommand{\qq}{{\frac{|\Psi_1|^2}{m_1}}}
\newcommand{\ww}{{\frac{|\Psi_2|^2}{m_2}}}
Interaction of vortices is an interesting question even in the 
simplest one-component Ginzburg-Landau (GL) Model \cite{aaa}. 
Its character in this model depends on the GL parameter 
$\kappa=\lambda/\xi$. In the type-II case (when 
$\kappa>1/\sqrt{2}$) 
the vortices repel each other. Another 
feature of the regime with $\kappa>1/\sqrt{2}$ 
is that a multiquanta vortex is unstable 
against decay into one-flux quanta vortices.
An invasion of vortices in a type-II
superconductor manifests itself 
as a second order transition 
at  a critical value  of applied field $H_{c1}$. 
 In contrast, in  the   type-I case
($\kappa<1/\sqrt{2}$)
vortices attract each other. A system of two
of  these vortices is  unstable against a collapse into 
a single two-quanta vortex. 
In the type-I regime a vortex  can 
not form in an applied external field since 
it would require external field which exceeds thermodynamic
critical magnetic field. However an isolated 
type-I vortex is stable.  The  type-I  vortices
were   considered (along with type-II vortices) in cosmology 
\cite{Hindmarsh:1994re}, where they play important role and 
 should be produced  during a symmetry breaking 
phase transition in early universe \cite{Kibble:1976sj}. 
The special case $\kappa=1/\sqrt{2}$ is also very interesting, in particular
it was shown exactly that 
the vortices at this value of $\kappa$ do not interact 
\cite{Bogomolny:1975de}.

A general formula for interaction of two parallel  vortices with a 
separation $r$ is \cite{Kr}
\begin{equation}
U(r)=c_1^2(\lambda, \xi )K_{0}\left(\f{r}{\lambda}\right)-
c_2^{2}(\lambda, \xi)K_{0}\bigg(\f{r}{\sqrt{2}\xi}\bigg)
\label{u}
\end{equation}
where $K_{0}(r)$ is the Bessel function,  $c_{1,2}(\lambda, \xi)$
are slow functions of the GL parameter, and
 $c_1(\kappa=1/\sqrt{2})=c_2(\kappa=1/\sqrt{2})$.
The first term in (\ref{u}) is interaction due to 
particle currents. Indeed this interaction has the 
range of the London penetration $\lambda$. 
The second term is the attractive interaction due to overlap of cores.

However   experiments \cite{hubener}
show  that   in certain materials 
with $\kappa \approx 1/\sqrt{2}$  
vortices attract each other at a large scale
and repel each other at a shorter distance.
The nonmonotic interaction  
of vortices was reproduced in a modified one-gap GL  functional
with additional terms  in the regime 
$\kappa \approx 1/\sqrt{2}$ \cite{teor,teor2}.
Besides that we should mention a nonmonotonic
interaction which takes place in layered systems:
 there a van der Waals - type of
 attraction can be  produced by thermal fluctions 
or disorder \cite{attr2}.

Below we consider a  simplest two-gap model, without 
addition of the extra terms
of the type discussed in context of single-gap model in \cite{teor,teor2},  and 
without restricting discussion to the case $\kappa \approx 1/\sqrt{2}$.
We show that two-gap GL model allows a novel type of vortices (the
``double-core" vortices) which have
  nonmonotonic interaction  in a wide range of parameters
 as their intrinsic feature.
The effective attraction of these vortices has  a principally 
different nature comparing to the effects 
considered in \cite{teor,teor2,attr2}. Moreover, we show that   two-gap
superconductors under certain conditions 
display even  more exotic magnetic properties which have no
counterparts in single-gap systems.

The  two-gap superconductor  \cite{mult,legg} (known
in particle physics as a two-Higgs doublet model \cite{tdlee}) 
in the simplest case can be described by the following GL functional:
\beg
&&
F = 
\frac{\hbar^2}{4m_1} \left| \left( \nabla +
i \f{2e}{\hbar c} {\bf A}\right) \Psi_1 \right|^2 + 
\frac{\hbar^2}{4m_2}  \left| \left( \nabla +
i \f{2e}{\hbar c} {\bf A}\right) \Psi_2 \right|^2 
 \nonumber \\
&&
+ { V} (|\Psi_{1,2}|^2)+ \eta [\Psi_1^*\Psi_2+\Psi_2^*\Psi_1] 
+ \frac{{\bf H}^2}{8\pi}
\la{act}
\eee
where $\Psi_\alpha = |\Psi_\alpha|e^{i \phi_\alpha}$ and $
{V} (|\Psi_{1,2}|^2)=\sum_{\alpha=1,2}-b_\alpha|\Psi_\alpha|^2+ 
\frac{c_\alpha}{2}|\Psi_\alpha|^4
$ and $\eta $ is a characteristic of the interband Josephson coupling strength
\cite{legg}.

Recently two-gap superconductivity was observed in
in   $Mg B_2$ \cite{mgb} and  in $NbSe_2$ \cite{2h}.  
Two-gap models appear also in the  theoretical studies
of   liquid metallic hydrogen, 
which should allow superconductivity 
of both electronic and protonic Cooper pairs \cite{ashc2}, while
in liquid metallic deuterium a deuteron superfluidity 
may coexist with  electronic superconductivity \cite{ashc2}. Besides 
that in superconductors there could 
coexist two types of pairing (e.g. $s$- and $p$-wave condensates).

One can rearrange variables in (\ref{act}) as follows  \cite{we,frac,kt}:
\beg
&F&=\f{\hbar^2}{4}\f{\f{|\Psi_1|^2}{m_1}\f{|\Psi_2|^2}{m_2}}{
 \f{|\Psi_1|^2}{m_1} 
+\f{|\Psi_2|^2}{m_2} 
} (\nabla ({\phi_1-\phi_2}))^2 +
 \f{4\hbar^2}{  \f{|\Psi_1|^2}{m_1} 
+\f{|\Psi_2|^2}{m_2} } \times
\nonumber \\
&&
\Biggl\{\f{|\Psi_1|^2}{4m_1}\nabla \phi_1+
\f{|\Psi_2|^2}{4m_2}\nabla \phi_2 
- \f{2e}{c}{\bf A}\Biggl\}^2\nonumber \\ &&
+ \f{{\bf H}^2}{8\pi} + 2\eta|\Psi_1 \Psi_2|\cos(\phi_1-\phi_2)
+{V} (|\Psi_{1,2}|^2)
\la{new01}
\eee
Let us consider first the case $\eta=0$.
The first term features a gauge invariant 
phase difference $\gamma_n=\phi_1-\phi_2$.
Since no coupling to gauge field enters this term, 
then any vortex with nontrivial winding number 
of $\gamma_n$ has  divergent 
energy  per unit length \cite{frac}.
In this paper we consider only vortices 
with finite energy per unit length. These are the 
vortices with 
phase windings satisfying the condition 
$\Delta \gamma_n\equiv\Delta(\phi_1-\phi_2) = 0$
where $\Delta \gamma_n$ denotes the gain 
in the phase $\gamma_n$ when we go around a vortex core.
In the case of nonzero $\eta$, one can observe that, when 
$\eta>0$ 
the free energy (\ref{new01}) is minimized 
by $\phi_1-\phi_2=\pi$, while when 
$\eta< 0$ we have $\phi_1-\phi_2=0$.
Thus when one considers  topological defects
with $\Delta \gamma_n=0$
the fourth term in  (\ref{new01}) 
is phase-independent and can be absorbed 
in the potential term. 
In this paper we want to discuss magnetic properties of
two gap system in the simplest and most transparent 
case and for this reason we  put $\eta=0$. As discussed 
above the low-temperature magnetic properties of the cases of zero and nonzero 
$\eta$ are essentially the same since in both cases only the vortices
characterized by $\Delta \gamma_n=0$ have finite energy.
An extension to a nonzero $\eta$
as well as to an addition of terms proportional to 
$|\Psi_1|^2|\Psi_2|^2$  is trivial. 
So { for the case when $\Delta \gamma_n = 0$},
the remaining nontrivial terms in  GL functional are:
\beg
&&F^{\gamma_n = const}=
 \f{4\hbar^2}{  \f{|\Psi_1|^2}{m_1} 
+\f{|\Psi_2|^2}{m_2} }
\Biggl\{\Biggl(\f{|\Psi_1|^2}{4m_1}+
\f{|\Psi_2|^2}{4m_2}\biggr)\nabla \gamma_c 
- \nonumber \\
 &&\f{2e}{c}{\bf A}\Biggl\}^2 
+ \f{{\bf H}^2}{8\pi}
-b_1|\Psi_1|^2+ 
\frac{c_1}{2}|\Psi_1|^4 -b_2|\Psi_2|^2+ 
\frac{c_2}{2}|\Psi_2|^4,
\la{new02}
\eee
where $\gamma_c=(\phi_1+\phi_2)$
The equation for supercurrent for (\ref{new02}) is (see e.g. \cite{we,frac,kt})
\beg 
{\bf J} &=&  
\frac{i \hbar e  }{2 m_1}
\left\{\Psi_1^*\nabla \Psi_1-
\Psi_1 \nabla \Psi_1^*\right\}
+\frac{i \hbar e }{2 m_2}
\{\Psi_2^*\nabla \Psi_2 \nonumber  \\
&& -\Psi_2 \nabla \Psi_2^*\} 
-\f{2 e^2}{c} \left( \f{|\Psi_1|^2}{m_1} 
+\f{|\Psi_2|^2}{m_2} \right){\bf A}. \ 
\label{AA}
\eee
As it was shown in \cite{frac}, in this model, the vortices 
characterized 
by the following phase gains: $\Delta\gamma_n = 0;
\Delta\gamma_c = 4\pi k$ carry $k$-quanta 
of magnetic flux.

The model
(\ref{new02}) possesses  three characteristic length scales:
two coherence lengths $\xi_\alpha=\hbar/\sqrt{4m_\alpha b_\alpha}$ determined by the effective potential
and the magnetic field penetration length: 
\be
\lambda=\frac{c}{\sqrt{8\pi} e}\left[ \f{|\Psi_1|^2}{m_1}+
\f{|\Psi_2|^2}{m_2}\right]^{-1/2}. 
\ee

 Let also  introduce a notation $\lambda_\alpha =
  c\sqrt{m_\alpha}/(\sqrt{8\pi}e{|\Psi_\alpha|})$
Below we also use notations for 
thermonynamic critical magnetic fields for
{\it individual} condensates  $H_{ct (\alpha)}=\Phi_0/(2\sqrt{2}\pi\xi_\alpha
 \lambda_\alpha)$. In discussion when 
 the individual  condensate $\Psi_1$ is of type-II
we denote the first and the second critical magnetic 
of isolated  $\Psi_1$ 
as follows: $H_{c1 (1)}=
\Phi_0/(4 \pi \lambda_1^2)[\log (\lambda_1/\xi_1) + 0.08]$
and $H_{c2 (1)} = \Phi_0/(2 \pi \xi_1^2)$.

Apparently in the cases when 
$\xi_1 \approx \xi_2 >> \lambda$ and when $\xi_1\approx \xi_2 << \lambda$, 
magnetic properties of the two-gap superconductor
parallel those of  single-gap type-I and type-II superconductors
correspondingly. However in a general case, when one 
of the condensates is of type I and another is of type II,
 a two-gap system 
in exernal field has much richer phase diagram and
allows many types of different transitions 
in external field. 

In the case when 
$\xi_1 << \lambda_1$ and $\xi_2 >> \lambda_2$, and
when thermodynamic critical 
magnetic field $H_{ct(2)}$ for the type-I condensate $\Psi_2$ is much higher than 
$H_{c2(1)}$ for the type-II condensate $\Psi_1$, 
the system  undergoes a
first order transition from a  normal state 
immediately into a two-gap superconducting
state. This is because, in this case, at fields higher than $H_{ct(2)}$ 
the system does not allow nontrivial solution of the 
linearized GL equation for $\Psi_1$, while, when 
the field is lowered below $H_{ct(2)}$, there appears 
a transition immediately into 
two-gap superconducting state. In this state the magnetic 
field is screened in bulk of a sample largely due to
surface current $\Psi_2$  and 
nothing can preclude the appearance of the second
condensate $\Psi_1$ in the bulk of the sample
 even if the applied field is much larger
than $H_{c2(1)}$.  

Consider now  the case when 
$\lambda_1/\xi_1> 1/\sqrt{2};\lambda_2/\xi_2>1/\sqrt{2}; \xi_1<<\xi_2;
\lambda=(1/\lambda_1^2 +1/\lambda_2^2)^{-1/2}< \xi_2/\sqrt{2}$.
In this case we have a two-gap system made up of two type-II condensates,
however, somewhat counterintutively, the magnetic properties of this system are 
principally different from a type-II superconductor.
One can observe that  the resulting penetration length 
in this system can be smaller than one of the coherence lengths 
$\lambda< \xi_2/\sqrt{2}$, since both condensates contribute 
to $\lambda$, so a vortex in this system 
will consist of an inner core of characteristic size $\xi_1$,
a flux tube of size $\lambda$, and an extended ``outer" core 
of a characteristic size of $\xi_2$ larger than $\lambda$ \cite{comment}.
It means that, albeit at length scales
smaller than $\lambda$ the interaction of two vortices
is repulsive, at the same time the extended core 
gives to such a vortex an effective attraction
$U \propto K_{0}(r/\sqrt{2}\xi_2)$ for $r>\lambda$ 
in a range of $\sqrt{2} \xi$.
Thus in contrast to an ordinary type-II superconductor, there 
exists  a range of  parameters where the phase transition
 at $H_{c1}$ in this system 
is of the first order. This is because due to attractive interaction 
the energy of a lattice made up of $n$ vortices 
(with the lattice spacing determined by the minimum of the 
nonmonotonic interaction potential) has lower energy 
than $n$ isolated vortices. Correspondingly the
magnetization of such a superconductor should change
abruptly by a formation of a vortex lattice, instead of a gradual
invasion of isolated vortices. Also for the same reason  in a
 decreasing applied field the vortex lattice 
should survive even in fields lower than the critical 
field  where a formation of an isolated vortex becomes
energetically favored.  Another consequence of the 
effective attraction between vortices is that at a given low 
vortex density,
when average distance between the vortices is larger
than the distance determined by the  minimum of the effective potential,
the vortices will spontaneously  form lattice clusters
in between  areas of vortex-free  Meissner state.

Let us now turn to the case when 
$\lambda_1/\xi_1> 1/\sqrt{2};\lambda_2/\xi_2<1/\sqrt{2}; 
\xi_2 > \lambda$. 
Let us  consider a vortex in this case.
For a two-gap superconductor,
 vortex energy consists of  energies of  cores, 
kinetic energy of Meissner current  
and magnetic field energy. The magnetic field energy 
and the kinetic energy of the screening current is given by
the standard expression: 
$F_{m}=(1/8\pi )\int d^3 x {\bf H}^2 + (1/8\pi )\int d^3 x \lambda_{eff}^2 ({ \rm curl} {\bf H})^2$.
Here we note that in the present case  $|\Psi_2({\bf x})|^2$
varies slowly over the London penetration length $\lambda$. 
Magnetic field is screened at a distance 
from the core which is smaller than $\xi_2$. That means 
that only depleted density of Cooper pairs of condensate $\Psi_2$
participates  in the screening of magnetic field.
Thus one can not use the London penetration length $\lambda$
but should introduce an effective penetration length 
$\lambda_{eff} = [1/\lambda_1^2+1/{\tilde\lambda_2^2({\bf x})}]^{-1/2}$
where ${\tilde\lambda_2}({\bf x})=  c\sqrt{m_2}/( \sqrt{8\pi} e{|\Psi_2 ({\bf x})|})
 > \lambda_2 = c\sqrt{m_2}/( \sqrt{8\pi} e{|\bar{\Psi}_2|})$, where $|\bar{\Psi}_2|$
is the average value of the order parameter modulus.
In the case  when $\lambda_1 << \xi_2$, we have
 $\lambda_{eff}\approx \lambda_1$
which corresponds to the situation when
magnetic field is screened mostly by condensate 
$\Psi_1$. 
On the other hand in the case when $\xi_2 >> \lambda_1>>\lambda_2>>\xi_1$
the magnetic field is screened at a chracteristic length 
scale of Pippard penetration length  $\lambda=\lambda_2^P \approx (\lambda_2^2 \xi_2)^{1/3}$
of the condensate $\Psi_2$.
This expression is valid when $\lambda_2^P<\lambda_1$.
In the above expression for $F_m$ 
we   cut off integrals at the distance  $\xi_1$
from the center of the core in order to obtain an estimate 
of the vortex energy  which has a logarithmic accuracy.
With it the  energy  per
unit length of a one-flux-quanta vortex is
\be
{\cal E} \approx \left(\f{ \Phi_0}{4\pi\lambda_{eff}}\right)^2 \log \f{\lambda_{eff}}{\xi_1} + {\cal V}_{c1} +{\cal V}_{c2}
\label{ve}
\ee
Where ${\cal V}_{c\alpha }$ are the energies of the cores per unit length
which are of order of magnitude of 
$[core \ size]\times[condensation \ energy] $.
The estimate of the core energy can also be expressed as: 
\be {\cal V}_{c\alpha } \approx \f{\pi \xi_\alpha^2 H_{ct(\alpha)}^2}{8\pi}
=  \f{ \Phi_0^2}{8\pi}\f{e^2|\Psi_\alpha|^2}{c^2 m_\alpha}
=\f{1}{4} \left(\f{ \Phi_0}{4\pi}\right)^2\f{1}{\lambda^2_\alpha}
\ee
Apparently  ${\cal V}_{c1} +{\cal V}_{c2} \approx
 ({ \Phi_0}/{8\pi\lambda})^2$.


A straightforward calculation of the  field $H_{c1}^0$
characterizing 
when it is becoming energetically preferred
to let into a superconductor a {\it single} vortex gives:
\be
H_{c1}^0\approx \f{\Phi_0}{4\pi}\left[\f{1}{\lambda_{eff}^2} \log \f{\lambda_{eff}}{\xi_1} \right]
+\f{\Phi_0}{16\pi}\left[\f{1}{\lambda^2_1}+\f{1}{\lambda^2_2}\right]
\la{hc1}
\ee
The transition into mixed state should be of the first order since 
the extended core at length scale $\xi_2>\lambda_{eff}$ gives rise 
to attractive interaction of vortices so, a lattice of vortices with a spacing 
determined by the 
 minimum of the interaction potential is preferred over 
a system of widely separated  vortices. That is, the energy 
of a system of $n$ vortices  is minimized when vortices spontaneously form 
a lattice cluster  with overlapping outer cores.
We should observe that $H_{c1}^0$ in the present situation is 
larger than thermodynamic critical magnetic field of the 
type-I condensate which is 
 $H_{ct (2)}=\Phi_0/(2\sqrt{2}\pi\xi_2 \lambda_2)$. 
We stress that this  however does not mean 
that at the field (\ref{hc1}) the condensate $\Psi_2$ is completely 
depleted due to overlapping of outer cores. It is because 
when  the applied  field  is close to  $H_{c1}^0$, then
the field is mostly  screened by a supercurrent of the condensate $\Psi_1$,
which circulates along  sample's edge. Thus
 the vortex system is dilute, or
more precisely the intervortex distance is
only determined by the effective attraction.
{  So in contrast to original type-I and type-II behavior, the  two-gap 
superconductor in this regime displays a first order transition into an
inhomogeneous state consisting of  clusters of vortices, where the order parameter
$\Psi_2$ is depleted due to overlaps of outer cores. So the superconductivity 
in these vortex lattice ``droplets" is dominated by the order parameter $\Psi_1$
and is essentially a one-gap superconductivity. Since the vortex 
density depends on applied field while the
intervortex distance is determined 
by the nonmonotonic interaction potential, so,
besides these clusters of vortices,
there should be present domains of  two-gap superconductivity in the
vortexless Meissner state.}

One of the possible direction where this work can be extended is 
the interesting question for a numerical study if there
is a range of parameters when a triangular lattice is not the most
energetically preferred. 
Indeed, for a given density of vortices, at a certain applied field,
spacing of a square lattice may be closer 
 to length scale given by minimum of vortex 
interaction potential, than spacing of a
triangular lattice. However,  in a triangular lattice, 
albeit the intervortex distance is larger,  there are six instead of four
nearest neighbors. Apparently in
sufficiently  strong magnetic fields
the effective attraction at large length scale is irrelevant and  
vortices form a standard triangular lattice.  So this system  may
allow  transitions between lattices with different symmetries, when 
applyed magnetic field is varied.

We also would like to note that in the recent experiment
on $MgB_2$ \cite{eskildsen} (which is the strictly type-II
two-gap superconductor) indicates a large
disparity in coherence lengths. The coherence length in $\pi$-band
was found being $\sim$50nm which  is much larger than an estimate 
which one would obtain from a standard GL formula for a one-gap 
superconductor using the $H_{c2}$ value of $MgB_2$.
Such a behavior is completely  in agreement  with the model
for the composite  two-gap vortex
\cite{frac}. 

In conclusion we discussed magnetic properties of a simplest form 
of a two-gap GL model. We have shown that the model exhibits 
behavior which has no counterpart in a  standard one-gap
superconductors. 
The discussed features  do not allow to classify the two-band superconductor
in the considered regimes as type-II and indeed nor as type-I but 
it should legitimately be distinguished in a separate class.

\vskip 1cm
{\bf Note added} A more detailed and quantitative study of the 
question which we discuss in this article can be found 
in  E. Babaev, J.M.  Speight {\tt cond-mat/0411681}


\begin{thebibliography}{99}
\bibitem{comment}
In fact the actual penetration length 
 will be larger than the London
penetration length $\lambda$ since it will 
receive a correction due to depletion of 
$|\Psi_2|$ at a distance smaller than $\xi_2$
from the core. The effective penetration length 
$\lambda_{eff}$
in this vortex, depending on the GL parameters,
will have the range $\lambda< \lambda_{eff}<\lambda_1<\xi_2$
\end{thebibliography}
\end{document}